\documentstyle[aps,prd,floats,epsf]{revtex}

\newcommand\lsim{\mathrel{\rlap{\lower4pt\hbox{\hskip1pt$\sim$}}
     \raise1pt\hbox{$<$}}}
\newcommand\gsim{\mathrel{\rlap{\lower4pt\hbox{\hskip1pt$\sim$}}
     \raise1pt\hbox{$>$}}}
\newcommand\esim{\mathrel{\rlap{\raise2pt\hbox{\hskip0pt$\sim$}}
     \lower1pt\hbox{$-$}}}

\preprint{DAMTP-2001-14}
\date{March 5th, 2001}
\tighten
\begin{document}


\title{Supernova constraints on spatial variations of
the vacuum energy density}

\author{P.\ P.\ Avelino${}^{1,2}$\thanks{
Electronic address: pedro\,@\,astro.up.pt}
J.\ P.\ M.\ de Carvalho${}^{1,3}$\thanks{
Electronic address: mauricio\,@\,astro.up.pt}
and C.\ J.\ A.\ P.\ Martins${}^{1,4}$\thanks{
Electronic address: C.J.A.P.Martins\,@\,damtp.cam.ac.uk}}

\address{${}^1$ Centro de Astrof\'{\i}sica, Universidade do Porto\\
Rua das Estrelas s/n, 4150-762 Porto, Portugal}

\address{${}^2$ Dep. de F{\' \i}sica da Faculdade de Ci\^encias da
Univ. do Porto,\\ Rua do Campo Alegre 687, 4169-007 Porto, Portugal}

\address{${}^3$ Dep. de Matem\'atica Aplicada da Faculdade de Ci\^encias da
Univ. do Porto,\\ Rua das Taipas 135, 4050 Porto, Portugal}

\address{${}^4$ Department of Applied Mathematics and Theoretical Physics\\
Centre for Mathematical Sciences, University of Cambridge\\
Wilberforce Road, Cambridge CB3 0WA, U.K.}

\maketitle
\begin{abstract}
{
We consider a very simple toy model for a spatially varying `cosmological
constant', where we are inside a spherical bubble
(with a given set of cosmological parameters) that is surrounded by a larger
region where these parameters are different. This model
includes essential features of more realistic scenarios
with a minimum number of parameters. We calculate the luminosity
distance in the presence of spatial variations of the vacuum energy
density using linear perturbation theory and discuss
the use of type Ia supernovae to impose constraints on this type of
models. We find that presently available observations are only
constraining at very low redshifts, but also provide independent
confirmation that the high red-shift supernovae data does prefer a
relatively large positive cosmological constant.
}
\end{abstract}

\section{Introduction}
\label{secintro}

The Standard Cosmological Principle (SCP) states that the universe is
spatially homogeneous and isotropic on large scales. It is
one of the cornerstones of modern cosmology, in the sense that if it
were not true, then the standard results for the most basic
properties of the universe (such as its geometry, content, or age) would
not hold and would have to be re-worked from scratch,
quite possibly suffering rather drastic changes.
There is some supporting evidence for the SCP on scales very close
to the horizon \cite{lahav1,lahav2,lahav3}, though it can hardly be
called definitive.

However, the theoretical situation is not totally unambiguous either.
On one hand, it has been claimed \cite{barcla} that
the SCP doesn't follow from the Cosmic Microwave Background Radiation (CMBR)
(near) isotropy and the Copernican Principle, as is usually assumed.
This is important because in that case it follows that
the SCP cannot be assumed to hold based on the presently
available observational data. On the other hand, it is known that there
are ways \cite{our1,our2,our3} in which the SCP could be evaded that would
be difficult, or even impossible, for us to notice at the present time.

A simple example which the present authors have considered in the
past \cite{our1,our2,our3} is that of a late-time phase transition
producing wall-like defects which separate regions with different values
of the cosmological parameters, notably the matter and vacuum energy
densities (and hence also the Hubble constant).
Since in such models the differences in the cosmological
parameter values in the different regions tend to increase with cosmic
time after the phase transition that produced them,
and since cosmological observations necessarily look back at earlier times,
detecting such differences is not as simple as one might expect,
and hence constraints on these models turn out to be surprisingly mild.

In addition, cosmological observations in such models will be faced
with the same type of limitations as were pointed out in \cite{maor}
for the case of quintessence models (see also \cite{wang,tegm} for different 
perspectives). It should be noticed that although
the two types of models have rather different motivations, they will
be rather similar observationally, since in both cases
one is effectively dealing with an equation state of the universe which
varies as a function of the redshift. However, an important
additional feature which needs to be taken into account in models with
different domains is the temperature jump due to the relative velocity
of comoving observers on either side of the domain wall. Also, these 
models will not be isotropic and so, in general, results 
will vary as a function of the direction on the sky. This means that 
problems with systematic errors due to supernovae evolution or dust 
are less severe than in the context of quintessence models.

In this paper we will use the recent measurements of the luminosity
redshift relation using supernovae
out to $z\sim1$ to constrain this type of models.
We should emphasise at the outset that we will only be dealing with
a very simple `toy model', but we nevertheless believe that it
still includes crucial features of more realistic scenarios with
a minimum number of parameters.

Supernovae have long been recognised as a crucial step in the
cosmological distance ladder, and hence as a useful tool to
estimate cosmological parameters via the Hubble diagram.
Recent observations of type Ia supernovae by two
independent teams \cite{riess,garn,perl}
(the Supernovae Cosmology Project and
the High-$z$ Supernova Search Team) out to redshift $z\sim1$
provided, together with CMB data, some fairly strong evidence for a
recently started phase of acceleration in the local universe,
with the preferred vacuum and matter densities being $\Omega_\Lambda\sim0.7$
and $\Omega_m\sim0.3$. This conclusion is based on the observed faintness
of high-redshift supernovae, relative to their expected brightness in a
`standard' decelerating universe.
It should be emphasised that these measurements are local and can not
be extrapolated all the way to the horizon. Therefore they do not,
on their own, imply that we have entered an
inflationary phase \cite{our1,stv}.

In the following section we introduce our toy model and discuss its possible
shortcomings. We then study the evolution of linear perturbations in our model
in Sect. \ref{growth}, and derive an expression for
the luminosity distance as a function of the red-shift in
Sect. \ref{seclumdis}. Finally, Sect. \ref{results} contains a
thorough discussion of our results, and we conclude in Sect. \ref{conclusions}.

\section{The model}
\label{secmodel}

In a recent article \cite{our3} we have shown that large sub-horizon
inhomogeneities may be generated if a network of domain walls
permeates the universe, dividing it in domains with slightly
different values of the vacuum energy density and other
cosmological parameters. The typical size of
these regions is determined by the
dynamics of the network of domain walls and is expected to be
close to the horizon scale.
The necessary condition in order
for the model to be observationally viable is that the domain walls
are formed in a late-time phase transition.

This condition is required for two
different (though related) reasons. Firstly, the cosmological
parameters will be different in the different domains, and the
differences tend to increase with time, so they must not be
so big as to make it observationally obvious at the present time.
Secondly, the domain walls can themselves be cosmologically
important (or even disastrous), and in order to avoid this one
requires that their energy scale is sufficiently
low so that they do not contribute in a significant manner to the
CMB anisotropies.

Here we study a simplified model in which the universe is made
up of a spherically symmetric region (domain), which is surrounded
by another region with a
different vacuum energy density (which we will call
$V_-$ and $V_+$ respectively).
We assume that we live in the centre of the inner region.
Moreover, one assumes that the thin region separating the two domains
considered (domain wall) does not generate relevant CMB fluctuations.
This happens if the potential of the field is small enough at the origin.
We also require that the domain walls have no non-trivial dynamics,
which is a good approximation if friction is important \cite{ms1,ms2,ms3}.

The vacuum density will be parametrised by
\begin{equation}
\Omega_\Lambda \equiv \frac{\rho_\Lambda}{\rho_c},
\label{defol}
\end{equation}
where $\rho_c$ is the critical density,
and we define $\Delta \Omega_\Lambda$ as
\begin{equation}
\Delta \Omega_\Lambda (r) = \frac{\delta \rho_\Lambda(r)}{\rho_c} =
\frac{\rho_\Lambda(r) - V_-}{\rho_c}.
\label{defdol}
\end{equation}
where $\rho_\Lambda(r)$ is the vacuum energy density
at the point in question. Hence, this can have two possible
values:  $0$ if we are inside the inner region,
and $(V_+-V_-)/\rho_c$ in the outer domain.

We emphasise that this is a highly simplified model. In more realistic
models, domain walls will have a non-trivial
dynamics and the domains will have many different shapes. Furthermore we
are also not expected to be exactly at the centre of a given domain.
In general, such cases would have to be dealt with
numerically. We will analyze some of these
effects elsewhere\cite{accm}. However, this simplified model
still incorporates some of the crucial features of more realistic models
with a minimum number of new parameters (the red-shift of the domain wall
$z_*$ and the difference between the two vacuum energy densities
$\Delta \Omega_\Lambda$). In the next sections we will show how
high-redshift supernovae can be used to constrain combinations of these
parameters. We shall also discuss the validity of these results in the
context of more realistic models in which a network of domain walls is
present.

\section{Evolution of cosmological perturbations}
\label{growth}

In the conformal-Newtonian gauge, the line-element for a flat
Friedmann-Robertson-Walker background and scalar metric perturbations
can be written as
\begin{equation}
ds^2 = a^2(\eta) \left[(1+2 \Phi) c^2 d \eta^2 - (1-2 \Phi) (dr^2 +
r^2 d
\theta^2 + r^2 \sin^2 \theta d \phi^2) \right],
\label{metric}
\end{equation}
assuming that the anisotropic stresses are small. Here, $\Phi$ is the metric
perturbation, $c$ is the speed of light in vacuum, $a$ is the
scale factor, $\eta$ is the conformal time, and $r$,
$\theta$ and $\phi$ are spatial coordinates.

Given that the vacuum energy becomes dominant
only for recent epochs we shall be concerned with
the evolution of perturbations only in the matter-dominated era,
neglecting the contribution of the radiation component.
The evolution of the scale factor $a$ is governed by the
Friedmann equation\footnote{A dot denotes a derivative with respect to
conformal time, ${\cal H} ={\dot a} / a$, the index `0'
means that the quantities are to be evaluated at the present time and
we have taken $a_0 = 1$ and ${\cal H}_0 = 1$ (so that the conformal time is
measured in units of ${\cal H}^{-1}_0$).}
\begin{equation} {\cal H}^2  = {{\Omega}}_m^0 a^{-1} +
{\Omega}_\Lambda^0 a^2\, .
\label{friedmann}
\end{equation}
Note that the background matter and vacuum energy densities at an
arbitrary epoch can be written as
\begin{equation} {\Omega}_m=\frac{{{\Omega}}_m^0}{{{\Omega}}_m^0 +
{\Omega}_\Lambda^0 a^3}
\label{omt}
\end{equation}
and ${\Omega}_\Lambda = 1-{\Omega}_m$.

The linear evolution equation of the scalar perturbations is given by
(see for example \cite{mfb})
\begin{equation}
    \ddot \Phi + 3 {\cal H} \dot \Phi  +
[2 {\dot {\cal H}} + {\cal H}^2]\Phi = 4 \pi G a^2 \delta p =
- {3 \over 2} a^2 \Delta \Omega_\Lambda ,
     \label{scalar}
\end{equation}
where $\delta p$ is the pressure perturbation and
\begin{equation}
2 {\dot {\cal H}}  = - { {\Omega}}_m^0 a^{-1} +
2 {\Omega}_\Lambda^0 a^2.
\label{doth}
\end{equation}
Given that the source term in the outer domain
($\propto a^2 \Delta \Omega_\Lambda$) is
only important near the present time, we shall assume the
following initial conditions for eq.~(\ref{scalar}):
\begin{equation}
\Phi(0)=0, \qquad {\dot \Phi}(0)=0,
\end{equation}
both in the inner and the outer regions. We note that given that the
source term
is absent in the inner region, the metric perturbation is always zero there.

The density perturbation, $\delta$, in the outer domain is given by\cite{mfb}
\begin{equation}
\delta \equiv {{\delta \rho} \over \rho_c}= 2 {{\Delta {\cal H}}
\over {\cal H}}  = {2 \over 3} {\cal H}^{-2} [\nabla^2 \Phi -3 {\cal
H} {\dot \Phi} -3 {\cal H}^2 \Phi],
\label{del}
\end{equation}
which simplifies to
\begin{equation}
\delta/2 = {{\Delta {\cal H}} \over {\cal H}}  = - ({\cal H}^{-1}
{\dot \Phi} +  \Phi),
\label{dels}
\end{equation}
given that $\nabla^2 \Phi = 0$ except at the domain walls.
The relationship between
$\delta$ and the fractional perturbation in the expansion rate follows directly
from the Friedmann equation.  It is straightforward to show that if the metric
perturbations are small the outer domain behaves
as having an effective $\Omega^{\rm eff}_\Lambda$ and $\Omega^{\rm eff}_m$
given by
\begin{equation}
\Omega^{\rm eff}_\Lambda = \left(1-2 {{\Delta {\cal H}} \over {\cal
H}}\right) \left(  { {\Omega}}_\Lambda + \Delta
\Omega_\Lambda \right)
\label{oleff}
\end{equation}
and similarly $\Omega^{\rm eff}_m = 1 - \Omega^{\rm eff}_\Lambda$.

\section{The luminosity distance}
\label{seclumdis}

We can now proceed by evaluating the luminosity distance relation in the
presence of spatial variations of the cosmological parameters.
Recall that the luminosity distance to a given source is given by:
\begin{equation}
d_L^2 = \frac{{\cal L}}{4 \pi {\cal F}}
\label{lumdis}
\end{equation}
where ${\cal L}$ is the luminosity of the source and ${\cal F}$
the measured flux.
It follows directly from the perturbed
flat FRW metric given in eqn.~(\ref{metric})
that, to first order in the metric perturbation, the luminosity distance,
$d_L$, to an object with comoving coordinate $r=r_1$
at a red-shift $z$ is given by:
\begin{equation}
d_L(z) \sim (1+z)[1- \Phi (\eta_0,0)] r_1 \sim  (1+z)[1- \Phi
(\eta_0,0)] c \int_{\eta(z)}^{\eta_0} [1+2 \Phi(\eta',|\eta_0 -\eta'|
{\bf n})] d \eta',
\label{lumdisz}
\end{equation}
where $\Phi (\eta',r')\equiv \Phi (\eta=\eta',r=r')$, $\eta_0$ is the
conformal time at the present time, ${\bf n}$ defines an arbitrary
direction on the sky and the observer is assumed to be at $r=0$.

Analogous corrections will also arise for the
Sachs-Wolfe effect. Here, in the presence of the scalar metric
perturbations defined by eqn. (\ref{scalar}) there is an additional shift
in the temperature of the source given by:
\begin{equation}
\frac{\Delta T}{T} = 2  \int_{\eta_e}^{\eta_0} d \eta {\dot \Phi}
(\eta,|\eta_0 -\eta| {\bf n}) + 2 \Phi (\eta_0,0) - 2 \Phi
(\eta_e,|\eta_0 -\eta_e| {\bf n}).
\label{sachsw}
\end{equation}
where $\eta_e$ is the conformal time when the light was emitted.
Within a given domain $\Delta T/T$ is
obviously zero. However, if the source is in the outer domain
there will be a temperature jump at the domain wall with
\begin{equation}
\frac{\Delta T}{T} = -2\Phi_+.
\label{dtot}
\end{equation}
where $\Phi_+$ is the value of $\Phi$ in the
outer region at the time when the light crossed the domain wall (here we are
assuming that $\Phi=0$ in the inner region).
This temperature shift is due to the relative velocity between comoving
observers on either side of the domain wall. This implies that for an observer
looking across the domain wall
the relation between the red-shift, $z$, and the scale factor,
$a$, has to be modified to
\begin{equation}
1+z = \frac{1}{a} - 2 \Phi_+\,.
\label{reds}
\end{equation}

These effects are a distinguishing characteristic of these type of models,
and could conceivably help distinguish them from
quintessence-type models, for example.

\section{Results and discussion}
\label{results}

We have verified the accuracy of our formalism by computing the luminosity
distance as a function of the red-shift of the source in two distinct ways.
In the first approach (case I, dashed line in
fig. 1) we assume that the
background universe has cosmological parameters $\Omega^0_\Lambda$,
$\Omega^0_m$ and ${\cal H}_0$. On top of this we introduce a perturbation
in the vacuum energy density parametrised by $\Delta \Omega^0_\Lambda$
and calculate $f(z)={\cal H}_0 d_L(z)/c$ using eqn.~(\ref{lumdisz}).
In the alternative approach (case II, solid line in fig. 1) we assume
that the cosmological parameters are $\Omega_m^{\rm eff}$,
$\Omega_\Lambda^{\rm eff}$ and ${\cal H}+\Delta{\cal H}$
(see eqns.~(\ref{dels}\ref{oleff})) with
no perturbation.

We have done the calculation for two distinct cosmological scenarios.
Model A has $\Omega^0_m=1$, $\Omega^0_\Lambda=0$ and
$\Delta \Omega^0_\Lambda=0.5$
while model B has $\Omega^0_m=0.3$, $\Omega^0_\Lambda=0.7$ and
$\Delta \Omega^0_\Lambda=0.2$. We can clearly see that the results obtained in
either case are in very good agreement for both models. In fig. 2 we
show the dependence of this
agreement on the value of $\Phi_0 \equiv \Phi(\eta_0)$ for model $B$ (here we
took $\Delta \Omega_\Lambda > 0$). We see that
for $|\Phi_0| \ll 1$ the relative agreement between the two methods for
calculating the luminosity distance
\begin{equation}
\epsilon=\frac{|d^I_L(\infty)-d^{II}_L(\infty)|}{d^I_L(\infty)}.
\label{relag}
\end{equation}
is nearly proportional to $|\Phi_0|^2$, as expected.

Having tested our method for calculating the luminosity distance we applied
it in the context of the model described in Sect. \ref{secmodel}  and used
type I supernovae in order to constrain the red-shift of the domain wall,
$z_*$, and the fluctuation in the values of the cosmological constant in the
outer region (parametrized by $\Delta \Omega^0_\Lambda$). The high-redshift
supernovae dataset of the Supernovae Cosmology Project was fit to the
FRW magnitude redshift relation:
\begin{equation}
m_B^{\rm eff} = {\cal M}_B + 5 \log D_L(z),
\label{magnitude}
\end{equation}
where $D_L(z) = {\cal H}_0 d_L(z)$,
${\cal M}_B$ is the ``Hubble-constant-free'' $B$-band absolute magnitude
at maximum of a supernovae with a stretch factor $s=1$, and $m_B^{\rm eff}$ is
the effective rest-frame $B$ magnitude corrected for the width-luminosity
relation. We assumed that $\Omega_m=0.3$ and $\Omega_\Lambda=0.7$ in the inner
region and we took ${\cal M}_B=-3.4$ neglecting the uncertainty
associated with the supernovae absolute magnitude calibration.
(We have also checked that
this particular choice does not affect our main results.)
We have estimated the parameters $z_*$ and
$\Delta \Omega_\Lambda^0$ using a $\chi^2$ statistical analysis.

Fig. 3 displays our results in the $z_*$ versus $\Delta \Omega^0_\Lambda$
plane. Note that for values of $|\Phi| \gsim 1$ our approach based on
linear perturbation theory ceases to be valid. The set of parameters for
which this happens is denoted as region I. The region of parameter space
that is allowed by the supernovae data (at $95 \%$ confidence) is
denoted as region II, while the observationally excluded region is
denoted as region III.

We clearly see that, as expected, if $z_*$ is small only a small value
of $\Delta \Omega_\Lambda^0$ is allowed.
However, as the value of $z_*$ increases, the constraints on the values of
$\Delta \Omega^0_\Lambda$ are much weaker, and there are essentially
no constraints beyond $z\sim0.7$. This happens essentially because
the importance of the cosmological constant is smaller in the past than
at the present time.

We also see that negative values of $\Delta \Omega_\Lambda^0$ are
excluded until significantly larger redshifts ($z\sim0.7$) than
positive ones (for which there are no constraints beyond $z\sim0.5$).
In fact the best fit model to the supernovae data has $z_* \sim 0.6$
and $\Delta \Omega_\Lambda^0 \sim 0.4$. This is a significant result---it
is an alternative (and perhaps intuitively clearer) way of saying that
the high red-shift supernovae data does favour a relatively large
positive cosmological constant.

\section{Conclusions}
\label{conclusions}

We have used type Ia supernovae data to constrain a simple toy model
for spatial variations of the cosmological constant. Specifically, the
model assumes that we are at the centre of a spherical region with a
given set of cosmological parameters which is enclosed by a
light domain wall and surrounded by another region where these parameters
are different. This aims to be a very simple mimic model for cosmological
models where the cosmological constant has space and/or time
variations (such as, eg quintessence models).

We have shown that the presently available supernovae data are only
constraining at very low red-shifts. Nevertheless, negative spatial
variations (meaning a present-day value of the cosmological constant
larger than the one at high red-shift) are much more constrained than
positive variations. In fact, our best-fit model turns out to have a
significantly larger vacuum energy density in the outer region than in
the inner one. This is a clear indication that a positive
vacuum energy density is favoured by the high red-shift data. Obviously
larger and deeper datasets will significantly improve these constraints.

Finally, we point out again that our analysis used a very simple toy
model. While we do believe that the model still captures the crucial
physics of the problem being discussed, it is clear that our analysis
can be improved by resorting to a numerical simulation of the
different domains. On the other hand, a careful discussion of the
effects of these spatial variations of cosmological parameters in the
CMB is also required. We shall return to these issues very shortly \cite{accm}.

\acknowledgements

We thank ``Funda{\c c}\~ao para a Ci\^encia e Tecnologia'' (FCT) for
financial support, and ``Centro de Astrof{\' \i}sica da Universidade do Porto''
(CAUP) for the facilities  provided.
C.\ M.\ is funded by FCT under  ``Programa PRAXIS XXI'' (grant no.\ 
FMRH/BPD/1600/2000).


\begin{figure}
\vbox{\centerline{
\epsfxsize=0.8\hsize\epsfbox{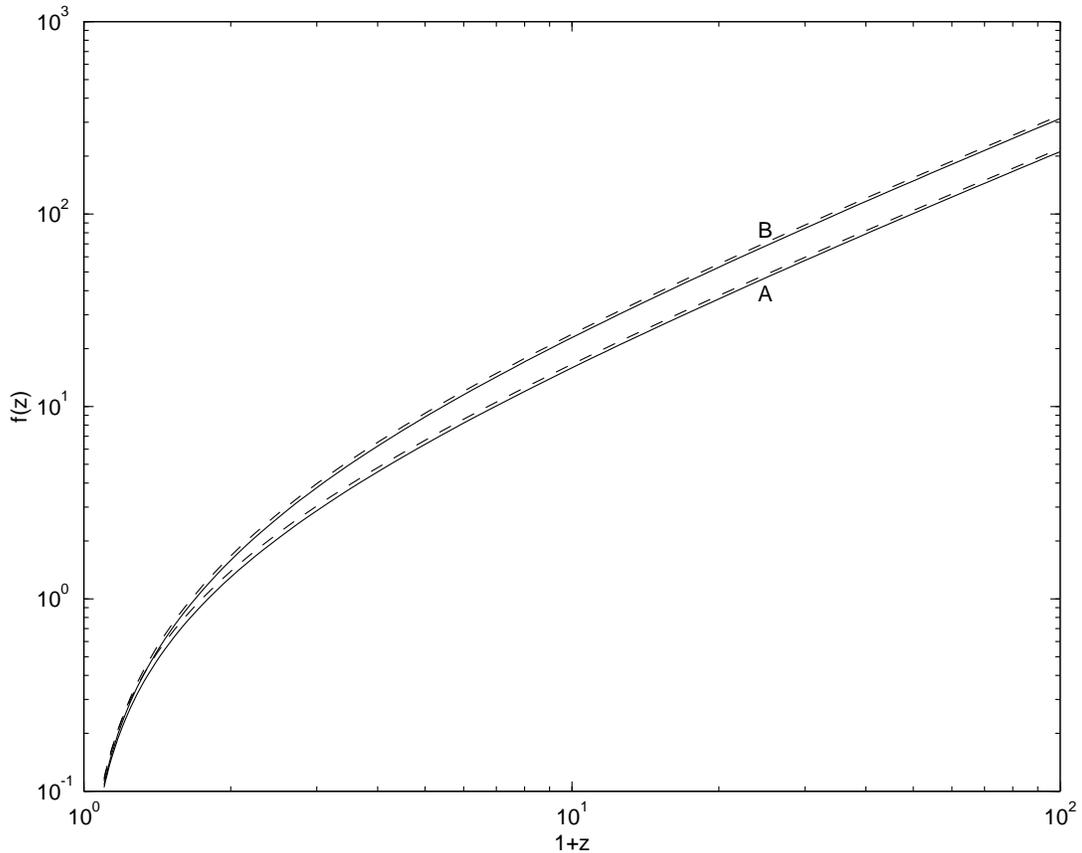}}
\vskip.2in}
\caption{Comparison between the luminosity distance, parametrized
by $f(z)=H_0 d_L(z)/c$, calculated
in two distinct ways. In case I (dashed line) we assume a
background universe with cosmological parameters $\Omega^0_\Lambda$,
$\Omega^0_m$ and ${\cal H}_0$ and a perturbation
in the vacuum energy density parametrised by $\Delta \Omega^0_\Lambda$.
In case II (solid line) we assume
that the cosmological parameters are $\Omega_m^{\rm eff}$,
$\Omega_\Lambda^{\rm eff}$ and ${\cal H}+\Delta{\cal H}$ with
no perturbation. Model A
has $\Omega^0_m=1$, $\Omega^0_\Lambda=0$ and $\Delta \Omega^0_\Lambda=0.5$
while model B has $\Omega^0_m=0.3$, $\Omega^0_\Lambda=0.7$ and
$\Delta \Omega^0_\Lambda=0.2$.}
\label{fig1}
\end{figure}

\begin{figure}
\vbox{\centerline{
\epsfxsize=0.8\hsize\epsfbox{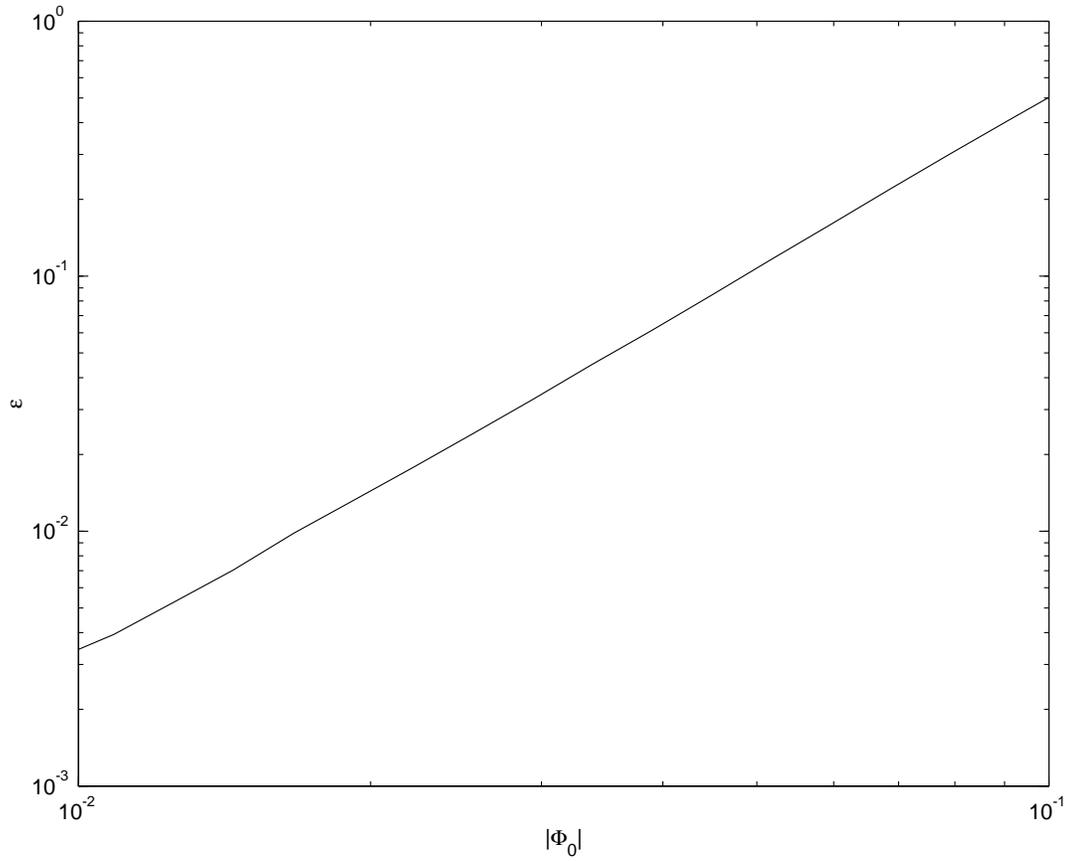}}
\vskip.2in}
\caption{Dependence the relative agreement between the two methods for
calculating the luminosity distance,$\epsilon$, on the value of
$|\Phi_0|$ for model $B$ (here we took $\Delta \Omega_\Lambda > 0$).
We see that $\epsilon$ is nearly proportional to $|\Phi_0|^2$, as expected.}
\label{fig2}
\end{figure}

\begin{figure}
\vbox{\centerline{
\epsfxsize=0.8\hsize\epsfbox{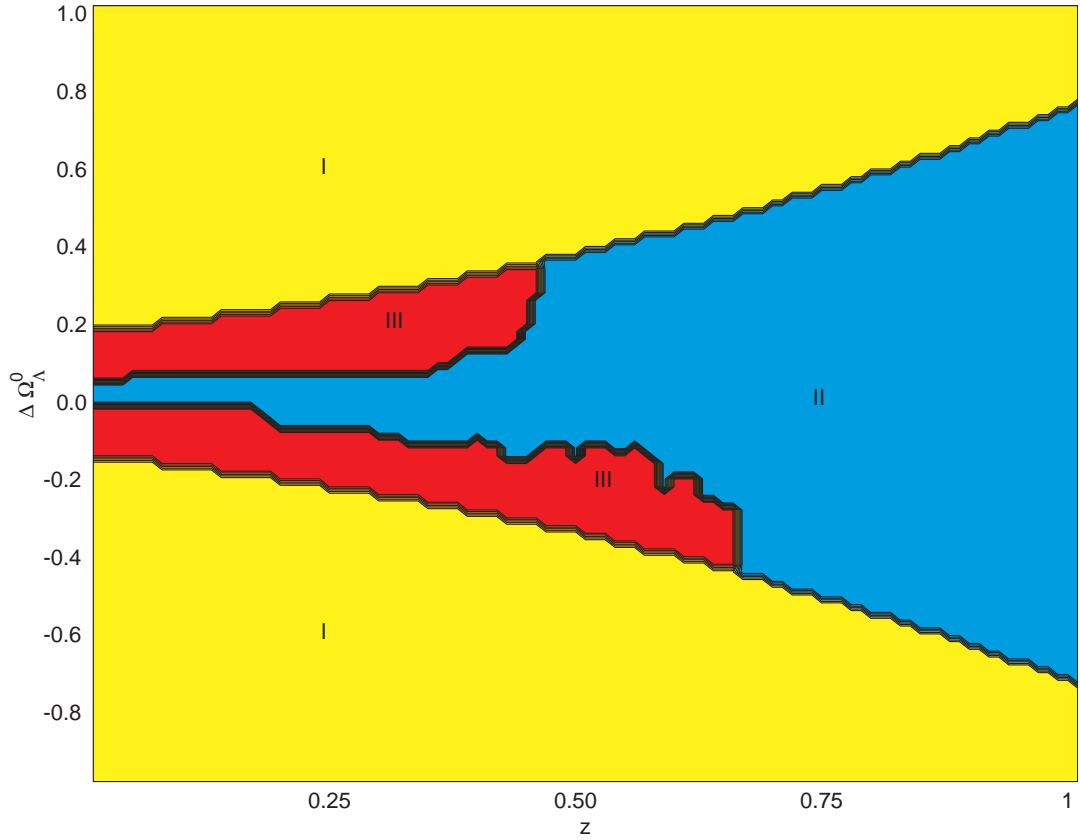}}
\vskip.2in}
\caption{The $95 \%$ confidence allowed region (region II) in the $z_*$
versus $\Delta \Omega^0_\Lambda$ plane. Region I is excluded since
our linear perturbation theory approach does not hold there, while
Region III is excluded by the supernovae data. Note the asymmetry between
the two observationally excluded regions.}
\label{fig3}
\end{figure}
\end{document}